\documentclass[proceedings]{JHEP3}
\PrHEP{PrHEP hep2001}
\conference{International Europhysics Conference on HEP} 
\usepackage{epsfig}

\newcommand{\leqsim}{\,\raisebox{-0.6ex}{$\buildrel < \over \sim$}\,}
\newcommand{\geqsim}{\,\raisebox{-0.6ex}{$\buildrel > \over \sim$}\,}
\newcommand{\unit}[1] {\mbox{\hspace{0.3em}\rm #1}}
\newcommand{\gev}{GeV$^2$}
\newcommand{\asmz}{\alpha_s(M_Z^2)}
\newcommand{\msbar}{\mbox{$\overline{\rm{MS}}$}\ }

\title{ZEUS NLOQCD fits}

\author{\speaker{Amanda Cooper-Sarkar}           \\  
        Address: NAPL,Keble Rd, OXFORD, UK \\     
        E-mail: \email{a.cooper-sarkar1@physics.ox.ac.uk}}

\abstract{ NLO QCD fits using the DGLAP formalism have been made to the high
precision ZEUS $e^+p$ reduced cross-section data and to fixed target structure
function data, in order to determine parton distribution functions and the
value of $\asmz$, taking full account of correlated systematic errors.
}

\begin{document}



The new high precision
ZEUS  data on the neutral current $e^+p$ reduced cross section~\cite{tuning}
are fit to the predictions of NLO QCD using the DGLAP equations,
in order to determine parton distribution functions and the
value of $\asmz$. The 
increased range ($6.3 \times 10^{-5} < x < 0.65$, $ 2.7 < Q^2 < 30000 GeV^2$)
and precision (systematic errors $\sim 3\%$, statistical errors $ \leqsim 1\%$,
for $2 < Q^2 < 800 GeV^2$) of the ZEUS data 
allow a much improved determination of the gluon and sea distributions 
compared to our previous work~\cite{zg94}. 
In recent years more emphasis has been
placed on estimating errors on extracted parton distributions.
We present parton distributions including full accounting for 
uncertainties from experimental correlated systematic errors. 
We are also
able to measure the value of $\asmz$ taking full account of the
correlations between the shape of the parton distribution functions and 
$\alpha_S$. Full details of the analysis are given in~\cite{forthcoming}.

This analysis is performed within the conventional
paradigm of leading twist, NLO QCD, with the
renormalisation and factorization scales chosen to be $Q^2$. 
The heavy quark production scheme is the
general mass variable flavour number scheme of Roberts and Thorne~\cite{hq}

The kinematics
of lepton hadron scattering is described in terms of the variables $Q^2$, the
invariant mass of the exchanged vector boson, Bjorken $x$, the fraction
of the momentum of the incoming nucleon taken by the struck quark (in the 
quark-parton model), and $y$ which measures the energy transfer between the
lepton and hadron systems.
The differential cross-section for the process is given in terms of the
structure functions by
\[
\frac {d^2\sigma } {dxdQ^2} =  \frac {2\pi\alpha^2} {Q^4 x}
\left[Y_+\,F_2(x,Q^2) - y^2 \,F_L(x,Q^2)
- Y_-\, xF_3(x,Q^2) \right],
\]
where $\displaystyle Y_\pm=1\pm(1-y)^2$. 
The structure functions $F_2$ and $xF_3$ are 
directly related to non-singlet and singlet quark distributions, and their
$Q^2$ dependence, or scaling violation, 
is predicted by pQCD. At $Q^2 \leqsim 1000$GeV$^2$ $F_2$ dominates the
charged lepton-hadron cross-section and for $x \leqsim 10^{-2}$ 
the gluon contribution
dominates the $Q^2$ evolution of $F_2$, such that ZEUS data provide 
crucial information
on quark and gluon distributions. (Schematically, $F_2 \sim xq$,
 $dF_2/dlnQ^2 \sim \alpha_s P_{qg} xg$).

We have performed a global fit of ZEUS and fixed target DIS data.
The fixed target data sets used are those with precision data  
for which full information on the correlated systematic errors is available
(NMC, E665, BCDMS  muon induced $F_2$ data on proton and deuterium targets
and CCFR $\nu,\bar{\nu}$ $xF_3$ data on an Fe target~\cite{hepdata}).
These data are used to gain information on the
valence quark distributions and the flavour composition of the sea, 
and to constrain the fits at
high $x$. However, our focus is on the additional information to be 
gained from 
the new precision ZEUS data, particularly on the 
gluon and quark densities at low $x$ and on the value of $\asmz$.

 In the standard fit the following cuts are made on
the ZEUS and the fixed target data: (i)~$W^2 > 20$~\gev\ to reduce the
sensitivity to target mass and higher 
twist contributions which become important at high $x$
and low $Q^2$; (ii)~$Q^2 > 2.5$~\gev\ to remain in the kinematic region where
perturbative QCD should be applicable.

The QCD predictions for the structure functions needed to construct the
reduced cross-section
are obtained by solving the DGLAP evolution 
equations at NLO in the \msbar\ scheme.
These equations yield the quark and gluon momentum distributions
(and thus the structure functions) at all values of $Q^2$ provided they
are input as functions of $x$ at some input scale $Q^2_0$.
The parton distribution functions (PDFs) for $u$ valence,  $d$ valence, 
total sea, gluon and the difference between the $d$ and $u$
contributions to the sea, are each parametrized  by the form 
\[
  p_1 x^{p_2} (1-x)^{p_3}( 1 + p_5 x)
\]
at $Q^2_0 = 7$GeV$^2$. Thus the flavour structure of the light quark sea 
allows for the violation of the Gottfried sum rule. We also impose
a suppression of the strange sea of a factor of 2 at $Q^2_0$,
consistent with neutrino induced dimuon data from CCFR. 

The parameters $p_1-p_5$ are constrained to impose the momentum sum-rule and 
the number sum-rules
on the valence distributions.
The gluon distribution has $p_5=0$, because non zero
values have minimal effect on the $\chi^2$ of the fit, and because this 
choice constrains the high $x$ gluon to be positive without the need 
for penalty $\chi^2$ terms.
There are 11 
free parameters in the standard fit when the strong coupling constant
 is fixed to $\asmz =  0.118$~\cite{lepalf},
and 12 free parameters when $\asmz$ is 
determined by the fit.

Full account has been taken of correlated experimental 
systematic errors as follows. 
The definition of the $\chi^2$ is
\[
\chi^2 = \sum_i \frac{\left[ F_i(p,s)-F_i(meas)) \right]^2}{(\sigma_{stat}^2+\sigma_{unc}^2)} + \sum_\lambda s^2_\lambda 
\]
where
\[
F_i(p,s) = \frac{F_i(NLOQCD)(p)} {\left[ 1 + 
\sum_{\lambda} s_{\lambda} \Delta^{sys}_{i\lambda}\right]}
\]
The latter equation modifies the NLOQCD prediction for the measured quantity
$F$ (structure function or reduced cross-section), which
is a function of the PDF parameters $p$, to include the effect of the 
correlated systematic errors. $\Delta^{sys}_{i\lambda}$ is the systematic 
error on data point $i$ due 
to source $\lambda$  and $s_\lambda$ are 
systematic error parameters. These systematic errors
are assumed to represent a one standard deviation error. The systematic 
error parameters are fixed to zero for the fit, 
but allowed 
to vary for error analysis, such that both the error matrices
\[
M_{jk}= \frac{1}{2}\frac{\delta^2 \chi^2}{\delta p_j \delta p_k}, 
C_{j\lambda} = \frac{1}{2}\frac{\delta^2 \chi^2}{\delta p_j \delta s_{\lambda}}
\]
are evaluated,
and the systematic covariance matrix is given by $V^{ps}= M^{-1} C C^T
M^{-1}$, so that the total covariance matrix is given by 
$V^{tot} = V^p +V^{ps}$, where $V^p=M^{-1}$.
This method is equivalent to offsetting each systematic parameter by $\pm 1$,
reminimizing and adding the resulting deviations, $\Delta p$,
from the parameters determined in the standard fit in quadrature~\cite{botnew}.
Thus it produces a conservative estimate of the effect of systematic errors.
Normalization errors are included as correlated systematic errors. 
In total 71 independent sources of systematic error
are included. 

A good description of
the structure function and reduced cross-section  
data over the whole range of $Q^2$ from 2.5 to $30000\unit{GeV}^2$
is obtained. The quality of the fit to the new ZEUS NC reduced cross-section 
and $F_2$ data is shown in~\cite{tuning}.
The quality of the fit to all the data sets may be judged from the
$\chi^2$. Adding the statistical and systematic errors in quadrature
gives a total $\chi^2$ per data point  of 0.95 for 1263 data points 
and 11 free parameters.

The parton distributions extracted from the fit are shown
in Fig.~\ref{fig:erd}.
\EPSFIGURE{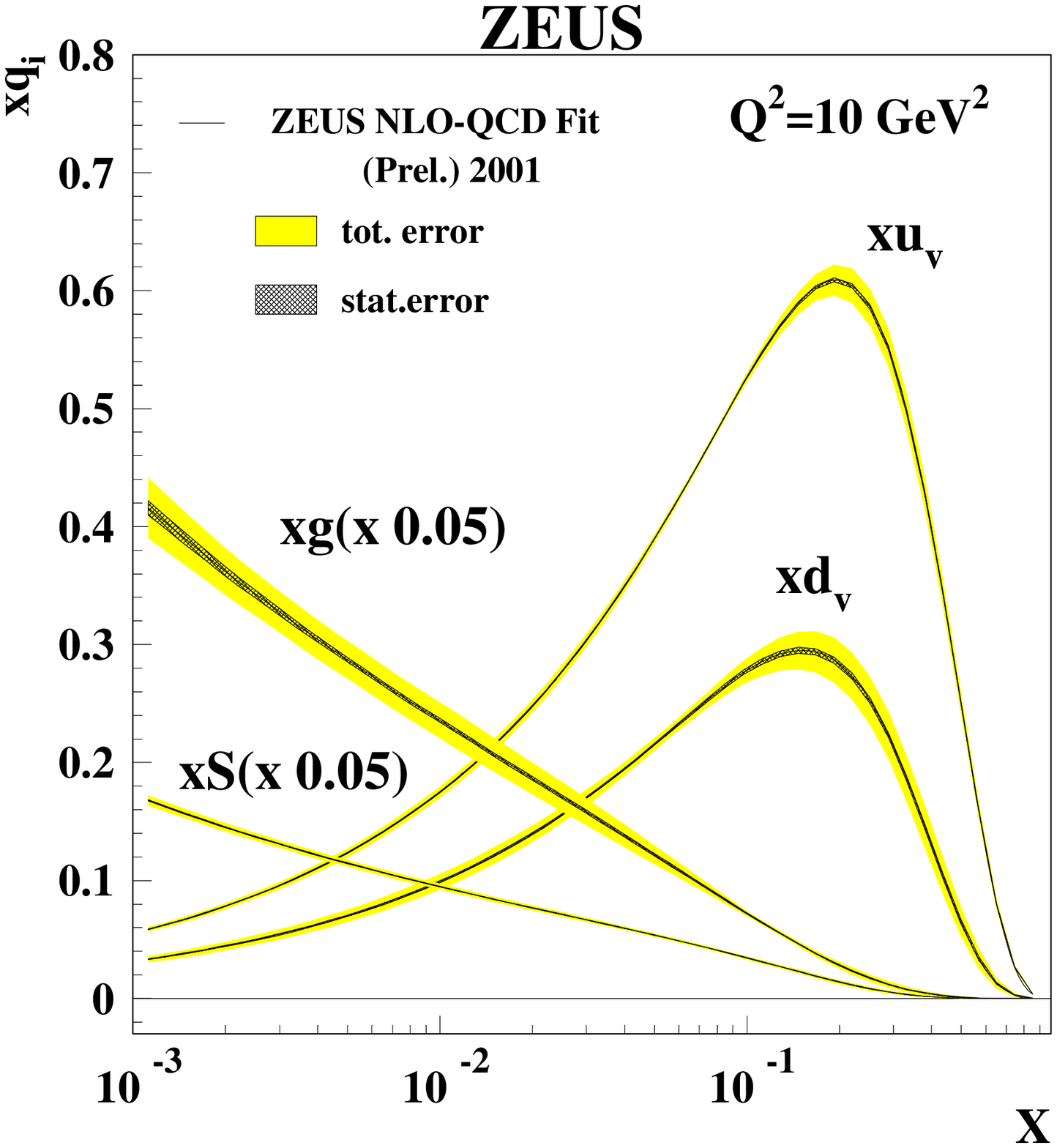,width=0.45\textwidth}
{The glue, sea, $u$ and $d$ valence distributions extracted from the 
standard ZEUS fit, at $Q^2=10$GeV$^2$, with the error bands resulting from
statistical, uncorrelated and correlated systematic errors.
\label{fig:erd}}
The ZEUS data are crucial
in determining the gluon and the sea distributions so we consider these in 
more detail. In Fig~\ref{fig:gbig} 
we show the extracted gluon distribution in several $Q^2$ bins. The
uncertainty in these distributions is further illustrated in Fig~\ref{fig:grat}
where the ratio of the error bands to the central value is shown.
In these figures the innermost error bands show the statistical and 
uncorrelated systematic error, the middle error bands show the total 
experimental error including correlated systematic errors and 
the outer error bands show the additional uncertainty coming from variation 
of the strong coupling constant 
$\asmz$, which is taken into account with full correlations by allowing 
$\asmz$ to be a parameter of the fit.
\DOUBLEFIGURE[b]{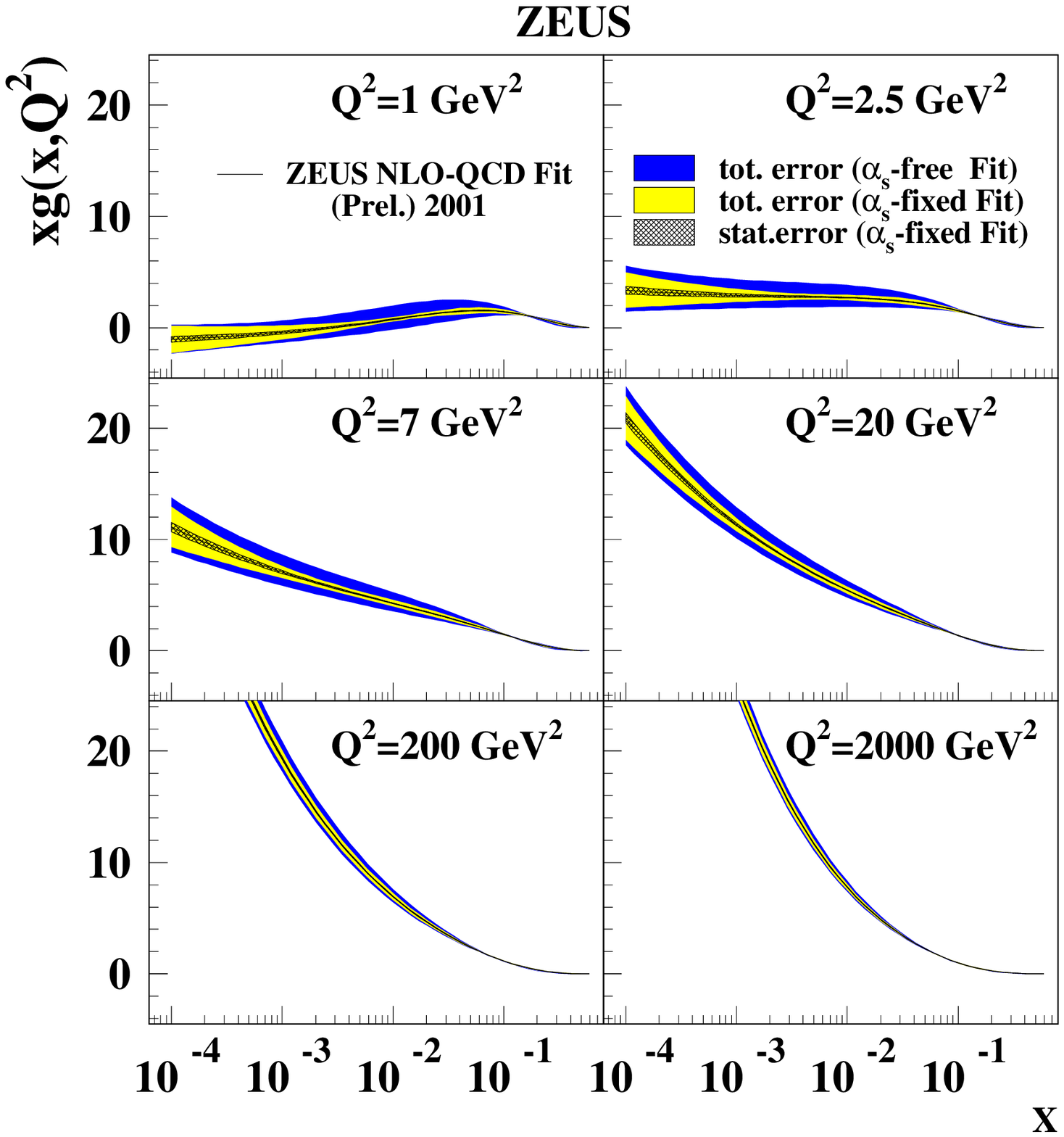,width=0.45\textwidth}
{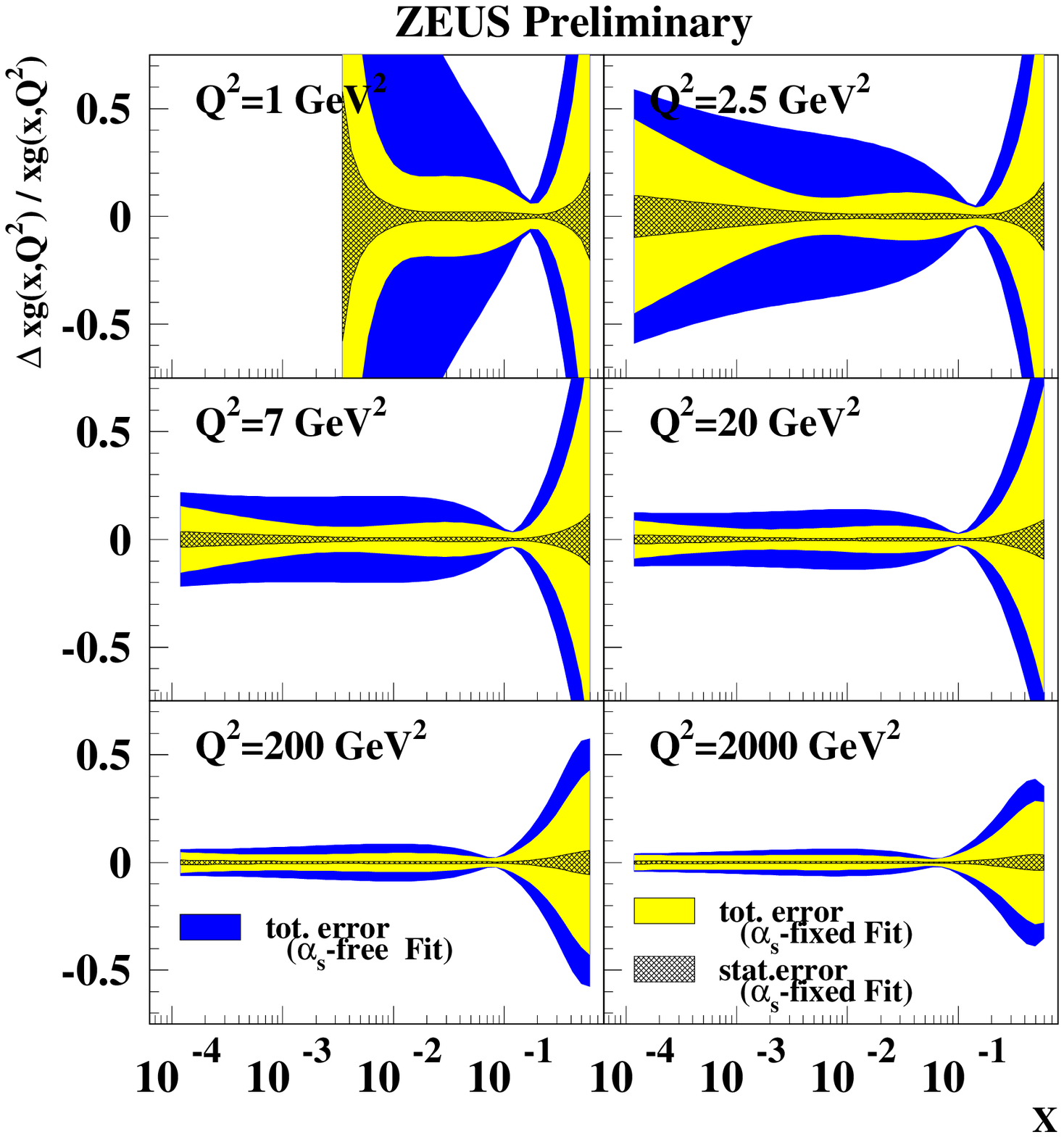,width=0.45\textwidth}
{The gluon distribution from the standard ZEUS NLOQCD fit.\label{fig:gbig}}
{The uncertainties on the gluon distribution from the standard 
ZEUS NLOQCD fit, shown as the ratio of the error bands to the central value.
\label{fig:grat}}

For $Q^2 \geqsim 5\unit{GeV}^2$, the gluon distribution rises steeply 
at low $x$ . It is well 
determined to within $\sim 10\%$ for $Q^2 > 20\unit{GeV}^2$,  
$10^{-4} < x < 10^ {-1}$, and its uncertainty
decreases as $Q^2$ evolves upwards. Considerable uncertainty remains for
$x > 0.1$, where one needs information from Tevatron jet studies 
or prompt photon data. At low $Q^2$ the gluon shape is flatter at low $x$, and
if the fit is extrapolated back to $Q^2 = 1\unit{GeV}^2$ the gluon shape
becomes valence like, with low $x$ values that are negative within 
errors. (This is also true of the structure function $F_L$ which is closely
related to the gluon.) 
The new precision ZEUS data collected in 96/97 shows this tendency
even more strongly than the previous 94/5 data~\cite{zg94}.

In  Fig~\ref{fig:sbig} 
we show the extracted sea distributions in several $Q^2$ bins. 
The uncertainty in these distributions
is is less than
$ \sim 5\%$ for $Q^2 \geqsim 2.5\unit{GeV}^2$ and  $10^{-4} < x < 10^ {-1}$. 
It can be seen that even at the
smallest $Q^2 \sim 1\unit{GeV}^2$ the sea distribution is still rising 
slightly at small $x$. In the same $Q^2, x$ region the gluon distribution
is falling and is consistent with zero. Thus the rise in the sea distribution 
cannot be driven by the gluon $g \rightarrow q\bar q$ splitting in this 
region.  
\EPSFIGURE{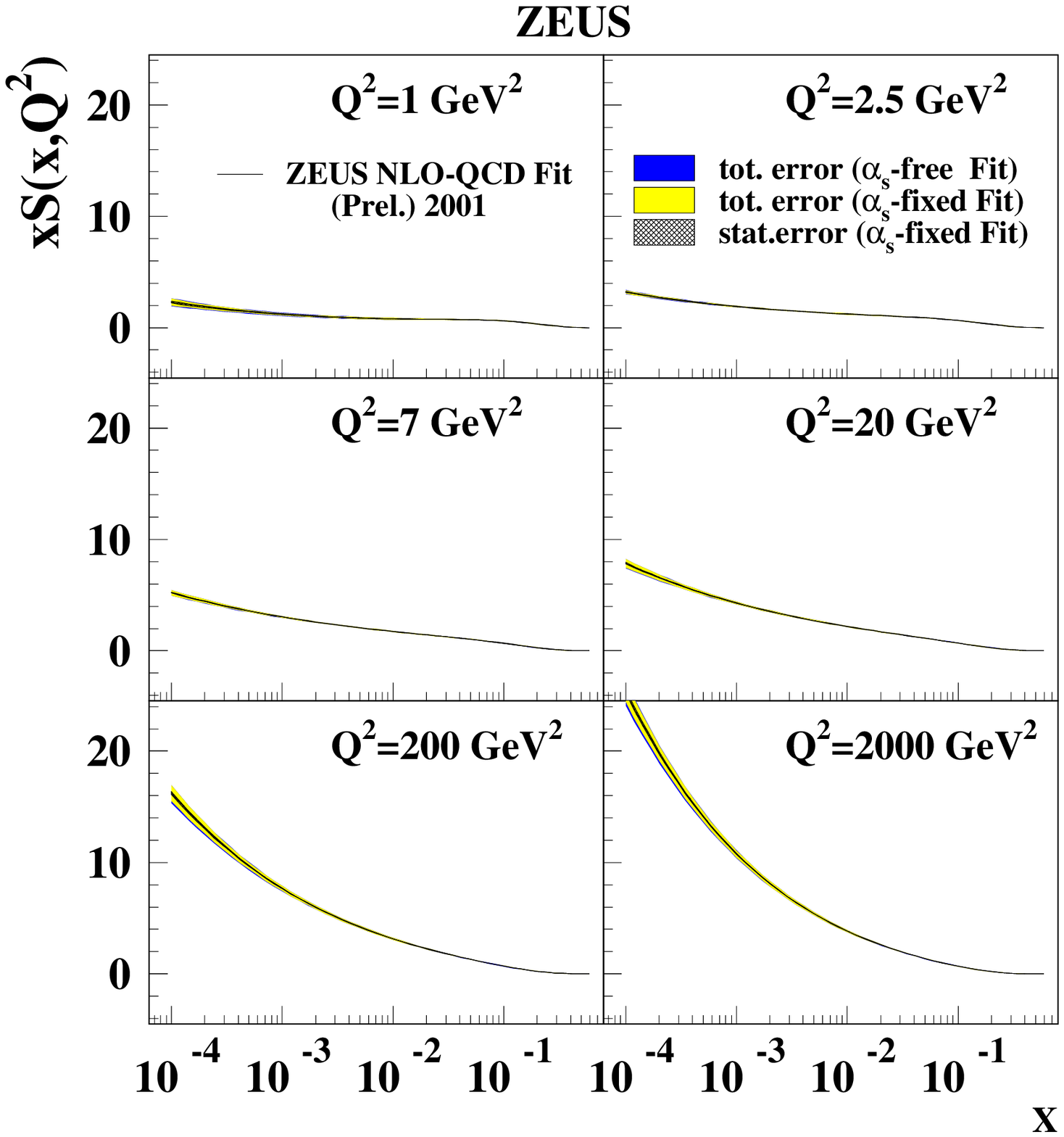,width=0.45\textwidth}
{The sea distribution from the standard ZEUS NLOQCD fit
\label{fig:sbig}}

We also make a fit using only ZEUS data, including the
charged current  $e^+p$ reduced cross-section data~\cite{ccpo} and
the neutral and charged current $e^-p$ data~\cite{ncel},~\cite{ccel}.
This high $Q^2$ data is very well described by the standard fit.
However now we use this data to constrain the valence distributions.
The $d_v$ valence distribution for this fit is shown in  
Fig~\ref{fig:dv_zeus} where the
$d_v$ distribution for the standard fit, including fixed target data, 
is shown for comparison. Clearly the standard fit
gives a more precise determination, but the high $Q^2$ ZEUS 
reduced cross-section
data is able to provide some constraint on the valence distributions.
We can see that 
ZEUS data prefer a larger $d$ valence distribution at high $x$ than the 
standard fit.
\EPSFIGURE{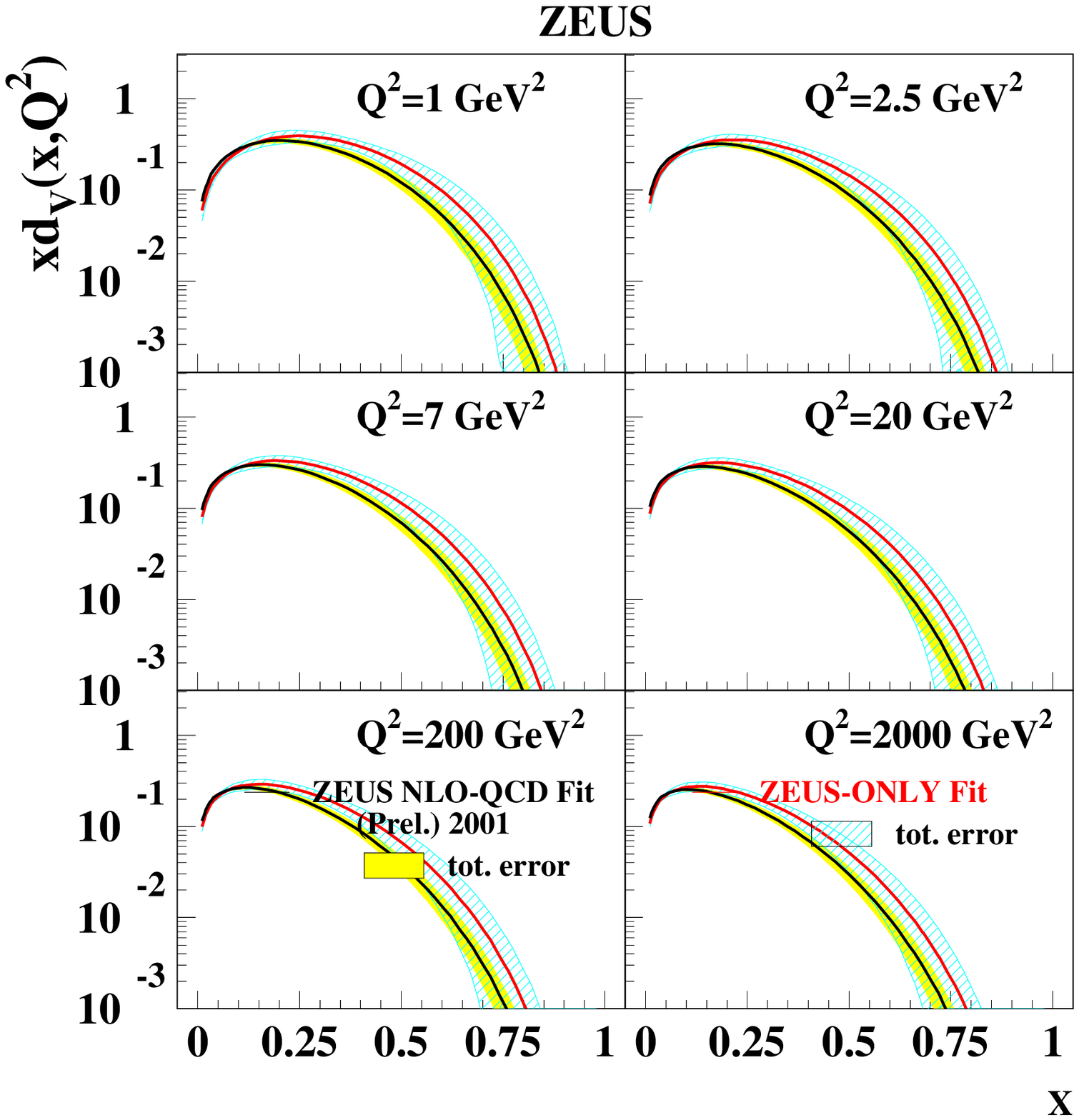,width=0.45\textwidth}
{The $xd_v$ distribution from the NLOQCD fit to ZEUS data alone.
\label{fig:dv_zeus}}

NLO DGLAP QCD fits give good descriptions
of $F_2$ data down to $Q^2$ values in the range $1-2\unit{GeV}^2$
but such fits assume
the validity of the NLO DGLAP QCD formalism 
even for low $Q^2$ (where $\alpha_S$ is becoming larger such
 that NNLO corrections
are increasingly important) 
and very low $x$
(where $ln(1/x)$ resummation terms could be important). 
The fits also ignore high density and non-perturbative effects
(see Refs.~\cite{cddrev} for a discussion of these effects).
To investigate if there is a technical low $Q^2$ limit to the NLO QCD fit
(in the sense that the fit fails to converge or gives a very bad $\chi^2$),
we extrapolate our standard fit into the region covered by the very low $Q^2$ 
ZEUS BPT~\cite{bpt97} data.
In Fig~\ref{fig:f2bpt} the main ZEUS data sample and the ZEUS BPT data are 
shown in very low $Q^2$ bins compared to the fit predictions.
\EPSFIGURE{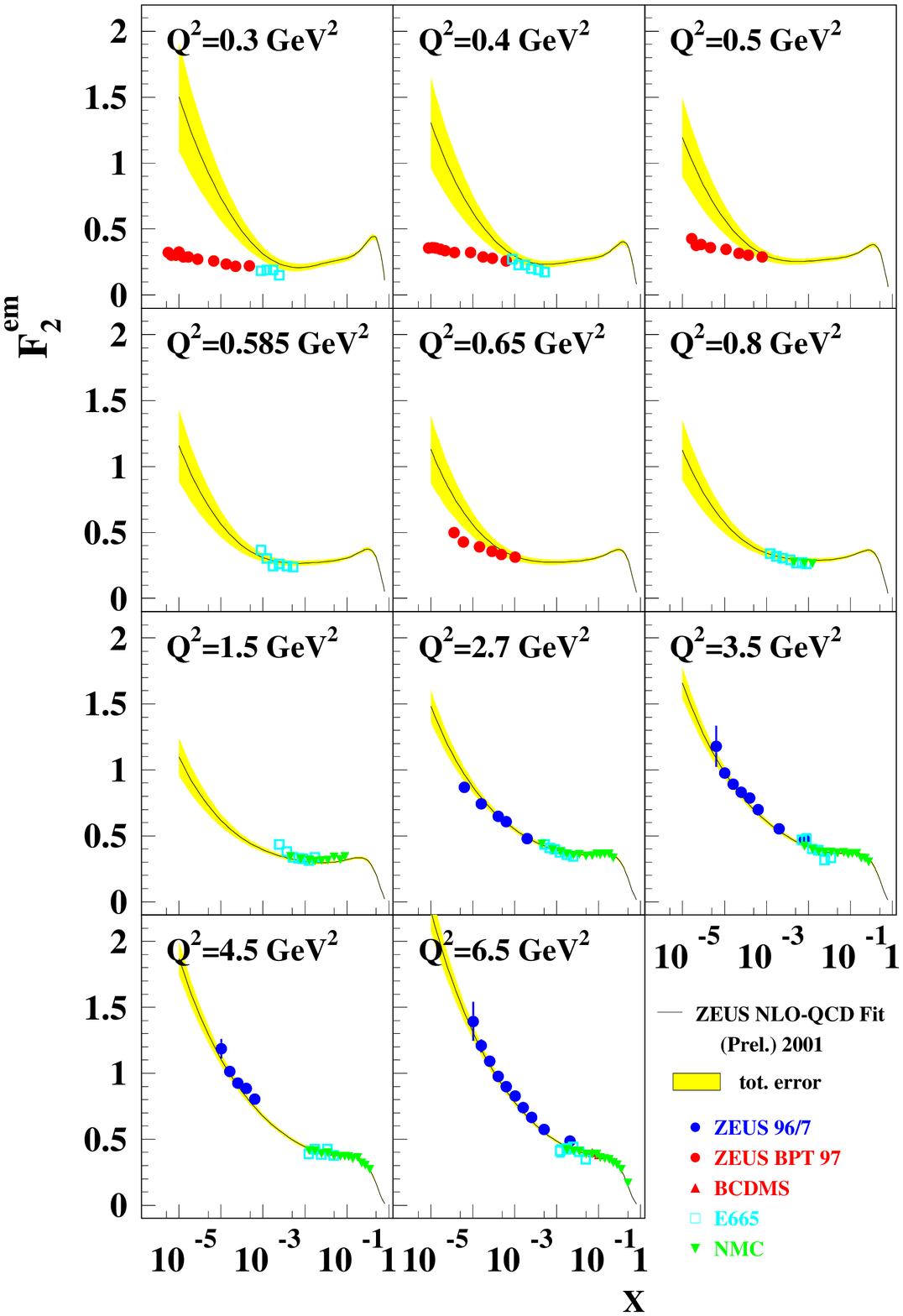,width=0.45\textwidth}
{$F_2$ data down to the very low $Q^2$ (including BPT data) compared
to the standard NLOQCD fit backward extrapolated.
\label{fig:f2bpt}}

We see that NLO DGLAP fit is unable to describe the 
very precise BPT data for 
$Q^2 \leqsim 0.65~\unit{GeV}^2$, even when the full error bands 
on the fit due to the
correlated systematic errors are included. This remains true even if these
data are included in the fit: they produce very poor fits, with 
$\chi^2$ per data point of $3.9$ or more. 
The predictions for $F_L$ for these low $Q^2$ values are also
significantly negative for $Q^2\leqsim 0.8~\unit{GeV}^2$ 
indicating that the NLO DGLAP formlism cannot be valid in this region.

The value of $\asmz$ is measured by allowing it to be a parameter of the fit.
We obtain,
\[
\asmz = 0.1172 \pm 0.0008(stat+uncorr) \pm 0.0054 (corr),
\]
where the first error represents statistical and uncorrelated systematic errors
and the second error represents the correlated systematic errors from all the
contributing experiments, including their normalization errors. The 
contribution of normalization errors is as large as half the
correlated systematic error.

The renormalization and factorization scales used in the
fit may be varied from
$Q^2/2 \rightarrow 2Q^2$,  as a way of 
estimating the importance of NNLO  and $ln(1/x)$ 
terms.
This produces variations in $\asmz$ of $\sim 0.004$. 

Within the
paradigm of NLO DGLAP the contribution of the correlated
systematic errors is the most significant source of uncertainty. 
Variation of analysis choices, such as the form of the parametrization at
$Q^2_0$, the value of $Q^2_0$, the minimum $Q^2$ of data entering the fit, 
do not produce a significant model error. 
A more significant choice is that of the heavy quark production scheme. 
Repeating the standard fit using the FFN scheme or the ZMVFN scheme produces
a variation in $\asmz$ of $\pm 0.001$.

The well known correlation of the value of $\asmz$ to the gluon shape is
accounted for by performing a simultaneous fit for $\asmz$ and the PDF
parameters. The largest 
uncertainty in the gluon shape is now at large $x$. Tevatron jet data 
suggests a larger gluon at high $x$ than that found in our standard 
fit, and if we use such data to constrain our fit we obtain
$\asmz \sim 0.121$,  a variation  within the quoted error.

Thus, when a full accounting of correlated systematic errors is
included, there is no contradiction between the  lower values of $\asmz$
originally determined from fixed target 
DIS data, e.g. $\asmz=0.113$ from BCDMS~\cite{mbb:marcv},
and the higher values favoured by LEP data $\asmz \sim 0.120$~\cite{lepalf}.


\bigskip
\par\noindent



\end{document}